\documentclass[twocolumn,showkeys,aps,prx, superscriptaddress]{revtex4-1}
\UseRawInputEncoding
\usepackage{graphicx}
\usepackage{pifont}
\usepackage[colorlinks=true,citecolor=blue]{hyperref}
\usepackage{graphicx}
\usepackage{amsmath, amssymb}
\usepackage{makecell}
\usepackage{multirow}
\usepackage{CJK}
\usepackage{mathrsfs}
\usepackage{bm}
\usepackage{amsmath}
\usepackage{dcolumn}
\usepackage{dsfont}
\usepackage{amssymb}
\usepackage{tabularx}
\usepackage{array}
\usepackage{float}
\usepackage{color}
\usepackage{mathrsfs}
\usepackage{extarrows}

\begin{document}

\title{Layer-Locked Anomalous Valley Hall Effect in \\Two-Dimensional A-Type Tetragonal Antiferromagnetic Insulator}

\author{San-Dong Guo}
\email{sandongyuwang@163.com}
\affiliation{School of Electronic Engineering, Xi'an University of Posts and Telecommunications, Xi'an 710121, China}

\author{Wei Xu}
\affiliation{State Key Laboratory of Surface Physics and Key Laboratory of Computational Physical Sciences (MOE), and Department of Physics,
Fudan University, Shanghai 200433, China}

\author{Yang Xue}
\affiliation{Department of Physics, East China University of Science and Technology, Shanghai 200237, China}

\author{Gangqiang Zhu}
\affiliation{School of Physics and Electronic Information, Shaanxi Normal University, Xi'an 716000, Shaanxi, China}

\author{Yee Sin Ang}
\email{yeesin\_ang@sutd.edu.sg}
\affiliation{Science, Mathematics and Technology (SMT) Cluster, Singapore University of Technology and Design, Singapore 487372}

\begin{abstract}
Antiferromagnetic (AFM) spintronics provides a route towards energy-efficient and ultrafast device applications. Achieving anomalous valley Hall effect (AVHE) in AFM  monolayers is thus of considerable interest for both fundamental condensed matter physics and device enginering. Here we propose a route to achieve AVHE in A-type AFM insulator composed of vertically-stacked monolayer quantum anomalous
Hall insulator (QAHI) with strain and electric field modulations. Uniaxial strain and electric field generate valley polarization and spin splitting, respectively. Using first-principles calculations, $\mathrm{Fe_2BrMgP}$ monolayer is predicted to be a prototype A-type AFM hosting \emph{valley-polarized quantum spin Hall insulator} (VQSHI) in which AVHE and quantum spin Hall effect (QSHE) are synergized in a single system. Our findings
reveal a route to achieve multiple Hall effects in 2D tetragonal AFM monolayers.

\end{abstract}

\maketitle

\textcolor{blue}{\textbf{Introduction.--}} Utilizing valley degree of
freedom to encode and process information, characterized as valleytronics, provides remarkable opportunities for developing next-generation minimized devices \cite{q1,q2,q3,q4, valley1, valley2, valley3}.
Valley refers to a local energy minimum/maximum in conduction/valence band,  where these energy extremes are robust against phonon and impurity scatterings due to the large separation in the momentum space\cite{q1}.
Recent proposals of valleytronics are mainly based on time-reversal-connected valleys, where
valley polarization is induced by an external field, dynamically or electrostatically \cite{q8-1,q8-2,q8-3,q9-1,q9-2,q9-3}.
Intrinsic valleytronics materials
with spontaneous valley polarization are more advantageous in terms of valley robustness, energy efficiency, and simplicity
in operation, which are beneficial for practical device applications.
Recently, the ferrovalley  (FV) semiconductor has been proposed \cite{q10}, which possesses spontaneous valley polarization induced by the combined effects of magnetic order and spin-orbit coupling (SOC).
Valley-dependent Berry curvature in FV materials leads to anomalous valley Hall effect (AVHE). FV materials thus offer an interesting platofrm to stuty valley-contrasting transport and Berry physics.

Achieving spontaneous valley polarization typically requires ferromagnetic (FM) system as a basic premise \cite{q10-1}.
Compared with ferromagnetism, antiferromagnetic (AFM) materials with zero magnetic moment are inherently robust to external magnetic perturbation, and possess ultrafast dynamics\cite{k1,k2},
thus offering enormous potential for both valleytronic and spintronic applications. However, spontaneous valley polarization in AFM materials is rarely
reported \cite{a1,a2,a3,a4,a5}, and the AVHE in AFM system is undesirable suppressed \cite{a2,a3}.
Furthermore, topological states based on valley-polarized quantum anomalous Hall insulator (VQAHI) has been recently proposed, in which valley polarization and quantum anomalous Hall effect (QAHE) are combined in one material.
Such system is particularly interesting dues to their compatibility with low-power electronics, spintronics, sensing, metrology and  quantum information processing applications \cite{q6,q7,q8}.
Several VQAHI systems have been theoretically proposed by constructing  complex heterostructure, layer-dependent proximity effects, and accurate regulation of strain and correlation strength \cite{q14,q15,q16,q17,q17-1,q17-2,q17-3,q17-4,re1,re2,re3,re4,re5,re6,re7}.

\begin{figure*}
  \includegraphics[width=18cm]{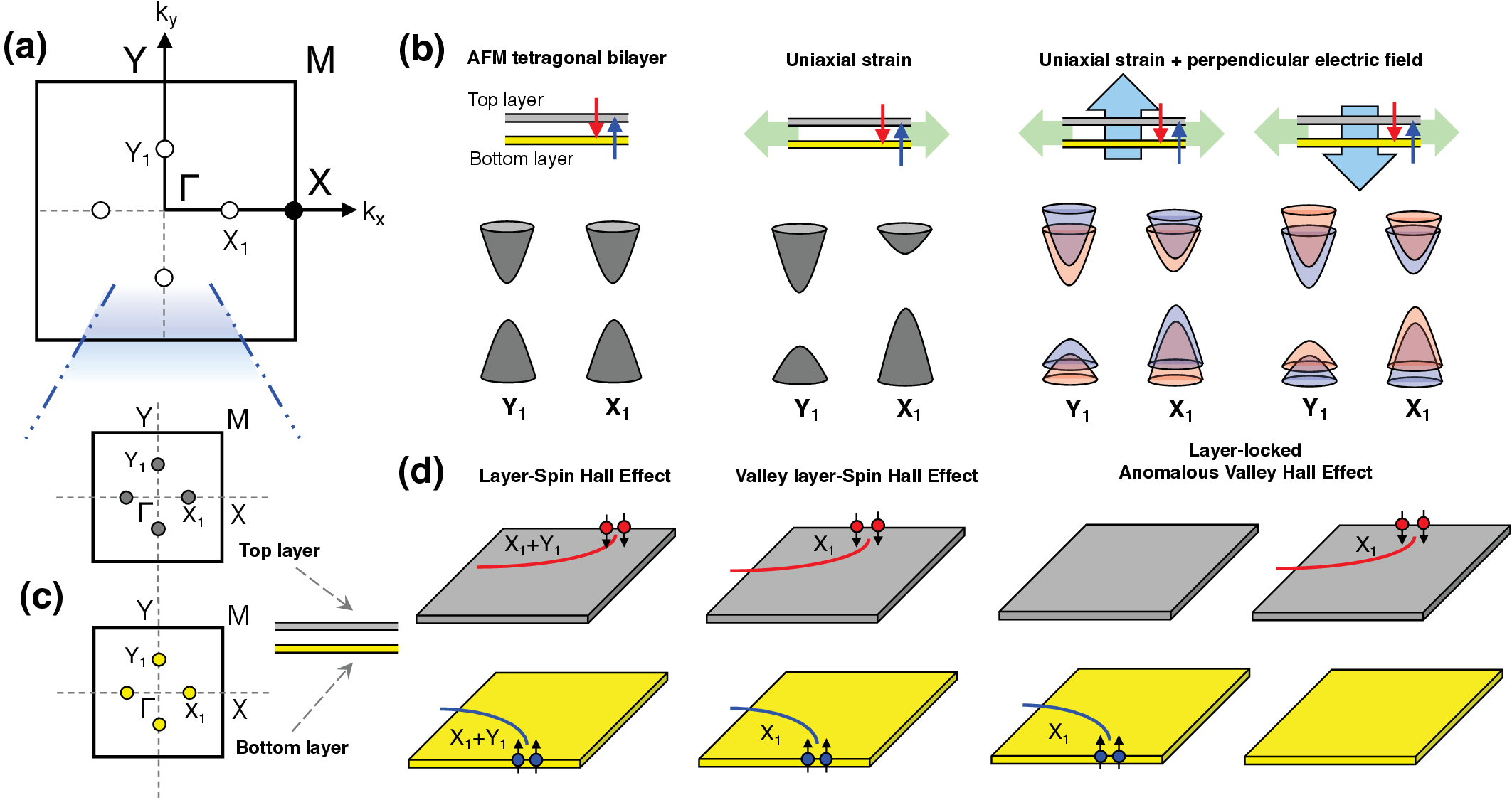}
\caption{\textbf{Concept of layer-locked anomalous valley Hall effect in 2D tetragonal lattice.}  (a) 2D tetragonal lattice possesses  equivalent valleys along $\Gamma$-X and $\Gamma$-Y lines with Berry curvature mainly occurring around $Y_1$ and $X_1$ valleys. (b) Applying uniaxial strain along $x$ direction makes the $Y_1$ and $X_1$ valleys become unequal but spin degeneracy is still preserved. Simultaneous application of uniaxial strain and out-of-plane electric field further breaks the spin degeneracy. Reversing electric field leads to opposite spin splitting at both $Y_1$ and $X_1$ valleys. The spin-up and spin-down channels are depicted in blue and red, respectively.
(c): Superposition of two tetragonal QAHIs   with equal but opposite magnetic moments (A-type AFM order) leads to spin-degenerate equivalent  $Y_1$ and $X_1$  valleys with net-zero Berry curvature in momentum space. However, the Berry curvatures for the spin-up and
spin-down channels are positive and negative, respectively, yielding nonzero layer-locked hidden Berry curvature in real space. (d) Showing the layer-spin hall effect, valley layer-spin hall effect, and layer-locked anomalous valley Hall effect.
}\label{fig1}
\end{figure*}

\begin{figure}
  \includegraphics[width=8cm]{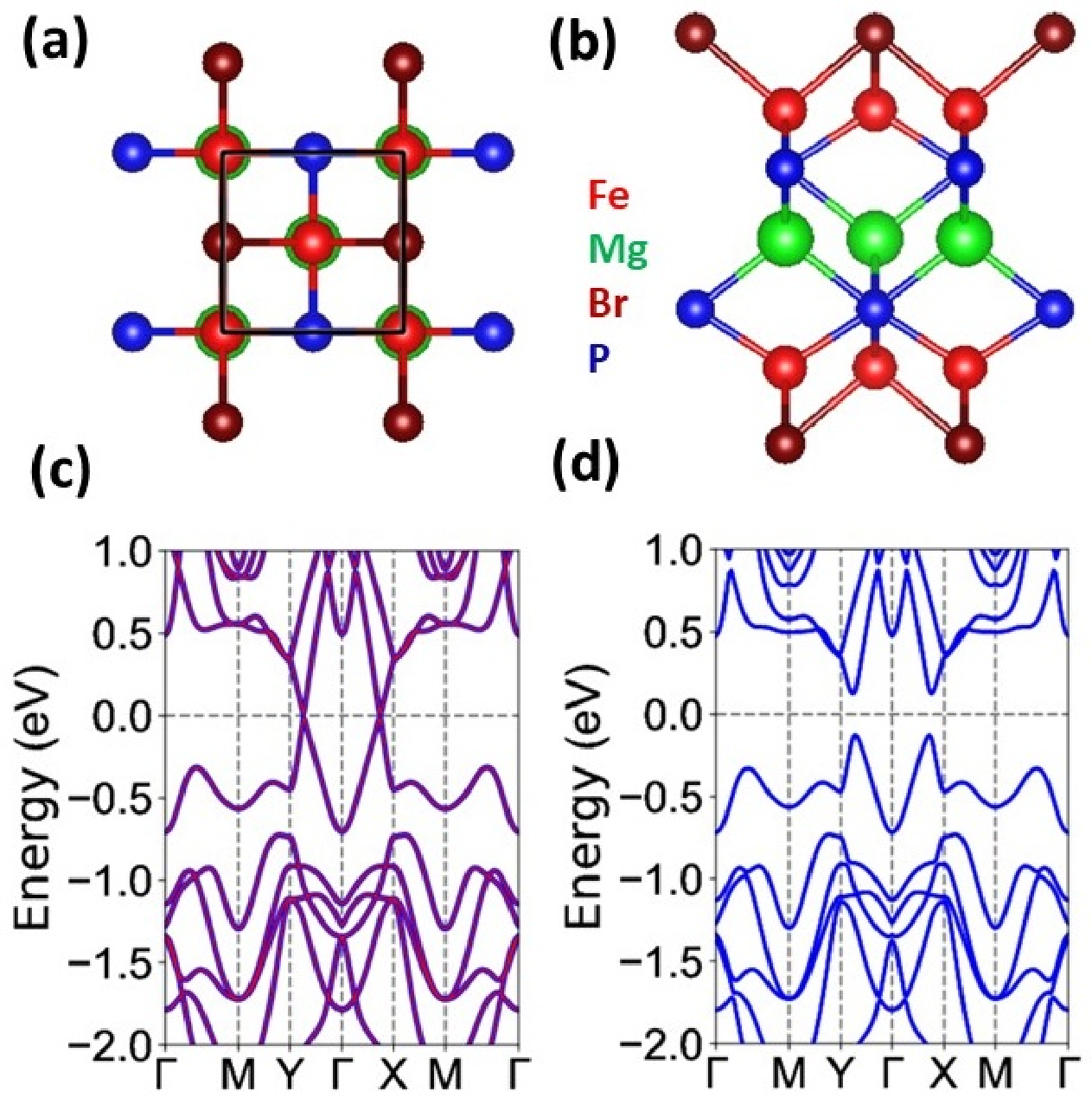}
  \caption{\textbf{Lattice and electronic structures of Fe$_2$BrMgP monolayer.} (a) and (b): top and side views of monolayer  $\mathrm{Fe_2BrMgP}$. Energy band structures of $\mathrm{Fe_2BrMgP}$ monolayer (c) without and (d) with SOC. In (c), the spin-up
and spin-down channels are depicted in blue and red.}\label{st}
\end{figure}

It should also be noted that previously reported valley-polarized AFM systems are mostly based on hexagonal symmetric lattice \cite{a1,a2,a3,a4,a5}.
Whether FV can be achieved in AFM system beyond hexagonal lattice symmetry remains an open question thus far. Can AVHE in AFM materials be achieved in system beyond hexagonal lattices?
Is it possible to achieve other valley-polarized topological state?
Here we propose a way to realize AVHE in an A-type tetragonal AFM insulator composed of vertically-stacked monolayered quantum anomalous
Hall insulator (QAHI) under both strain and electric field tuning. A peculiar valley-polarized quantum spin Hall insulator (VQSHI) can be achieved in the proposed systems, which is verified via first-principles in Fe$_2$BrMgP monolayer as a prototype VQSHI.
Our findings open up a previously unexplored concept of VQSHI in which valley-polarizaiton and quantum spin Hall effect (QSHE) are synergized in a single system.

\textcolor{blue}{\textbf{Achieving AVHE in AFM insulators.--}} Firstly, a two-dimensional (2D) tetragonal FM QAHI is used as the basic building block, which  has a layer of magnetic atoms and possesses equivalent valleys along $\Gamma$-X and $\Gamma$-Y lines in the first Brillouin zone (BZ) due to $C_4$ rotation symmetry [Fig.\ref{fig1} (a)]. The Berry curvatures mainly occur around the $Y_1$ and $X_1$ valleys with positive values.

To realize  AFM insulator, a superposition of two 2D tetragonal FM  QAHI with equal but opposite magnetic moments (A-type AFM order) is constructed [Fig.\ref{fig1} (b).], giving rise to spin degeneracy and equivalent  $Y_1$ and $X_1$ valleys. Here, we assume that the AFM system has a symmetry of a combination of inversion symmetry $\mathcal{P}$ and time-reversal symmetry $\mathcal{T}$ ($\mathcal{PT}$), which leads to
a spin-degenerate 2D system.
To induce valley polarization, a natural way is to destroy $C_4$ rotation symmetry via uniaxial strain  along the $x$ or $y$ direction [Fig.\ref{fig1} (b)]. However, the spin degeneracy is still maintained under uniaxial strain, which prohibits the AVHE. An out-of-plane  electric field $E_{\perp}$ is introduced to break the $\mathcal{PT}$ symmetry, which lifts the spin degeneracy of valleys (Fig.\ref{fig1} (b)). Such spin-degeneracy breaking is due to the layer-dependent electrostatic potential $\varpropto$ $eEd$ ($e$ and $d$ denote the electron charge and the layer distance) created by the out-of-plane electric field, which causes the spin-up and spin-down bands in different layers to stagger, leading to spin-splitting effect.
A similar mechanism can be found in electric potential difference antiferromagnetism \cite{epd}.
More interestingly, the spin orders at both $Y_1$ and $X_1$ valleys
can be reversed through reversing the direction of out-of-plane electric field [Fig.\ref{fig1} (b)], thus offering electrostatic field-tunable valley polarization.

The superposition of two tetragonal QAHIs leads to zero
Berry curvature $\Omega(k)$ in momentum space due to $\mathcal{PT}$ symmetry [Fig.\ref{fig1} (c)].
However, each layer breaks the $\mathcal{PT}$ symmetry \emph{individually}, which gives rise to the layer-locked \emph{hidden} Berry curvature, and the Berry curvatures for the spin-up and spin-down channels are positive and negative-valued, respectively. Such layer-locked hidden Berry
curvature leads to a peculiar layer-Hall effect not commonly found in other AVHE systems.

In the presence of a longitudinal in-plane electric field $E_{\parallel}$,  the Bloch carriers acquires an anomalous transverse velocity $v_{\bot}$$\sim$$E_{\parallel}\times\Omega(k)$ \cite{q4}. By
shifting the Fermi level in the valence band via hole
doping, various layer-spin Hall and layer-locked AVHE can occur [Fig.\ref{fig1} (d)]: (i) The spin-up and spin-down electrons from $Y_1$ and $X_1$ valleys accumulate along opposite sides of different layers in the case of 2D A-type tetragonal AFM system, resulting
in layer-spin hall effect; (ii) when a uniaxial strain is applied, the spin-up and spin-down electrons from only $X_1$  valley accumulate along the opposite sides of different layers, resulting in  valley layer-spin hall effect; (iii) the spin-up/spin-down electrons from only $X_1$  valley
accumulate along one side of bottom/top layer, resulting to the rarely explored \emph{layer-locked anomalous valley Hall effect}.

\textcolor{blue}{\textbf{Material realization.--}} Monolayer $\mathrm{Fe_2XY}$ (X=or$\neq$Y=Cl, Br and I) and $\mathrm{Li_2Fe_2XY}$ (X=or$\neq$Y=S, Se and Te) families\cite{fe,fe1,fe2,fe3,fe4,fe5} can be used as the basic building block. These monolayers are tetragonal QAHIs with equivalent valleys along $\Gamma$-X and $\Gamma$-Y lines  in the first Brillouin zone (BZ), and the extremes of Berry curvatures are
located at the $Y_1$ and $X_1$  valleys. Instead of employing the vertical stacking of two identical monolayers via van der Waals (vdW) heterostructure approach, we consider an intercalation architecture in which two identical monolayers are intercalated to form an `ultrathick'
$\mathrm{Fe_2XYP}$ monolayer (X=Br, Cl and I; Y=Mg and Be) \cite{x1}.
We use $\mathrm{Fe_2BrMgP}$ as a protype system to illustrate the concept of layer-locked AVHE in 2D tetragonal AFM. The first-principles calculation details are presented in the Supplementary Materials.

\begin{figure}
  \includegraphics[width=8cm]{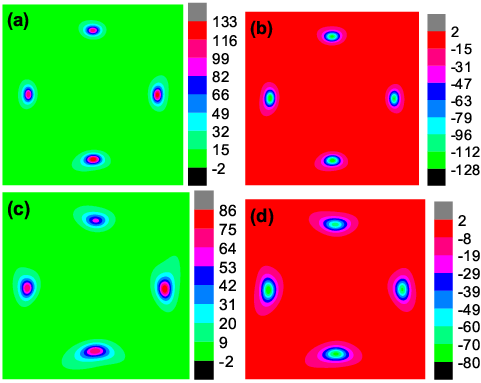}
\caption{\textbf{Layer-dependent Berry curvature of $\mathrm{Fe_2BrMgP}$.} The Berry curvatures are shown for the spin-up (a,c) and spin-down (b,d) channels for $a/a_0$=1.00 and $E$=0.00$\mathrm{V/{\AA}}$ (a,b), and for $a/a_0$=1.04 and $E$=0.02$\mathrm{V/{\AA}}$ (c,d).}\label{t1}
\end{figure}

\textcolor{blue}{\textbf{Lattice, magnetic and electronic properties.--}} $\mathrm{Fe_2BrMgP}$ monolayer is dynamically, mechanically, and thermally stable \cite{x1}. The crystal structures of $\mathrm{Fe_2BrMgP}$ monolayer  are plotted in Fig.\ref{st} (a) and (b), crystallizing in the  $P4/nmm$ space group (No.129).
The unit cell contains ten atoms with seven-atomic layer sequence of Br-Fe-P-Mg-P-Fe-Br.
The optimized equilibrium lattice constants are $a$=$b$=4.03 $\mathrm{{\AA}}$ by GGA+$U$ method.
To determine the ground state of $\mathrm{Fe_2BrMgP}$, we consider two magnetic configurations, including the intralayer FM and interlayer FM ordering (FM ordering), and intralayer FM and interlayer AFM ordering (A-type AFM ordering).
This A-type AFM ordering is predicted to be the ground state, and its energy is 54.86 meV per unit cell lower than that with
the FM ordering.
The different magnetic orientation can  affect the symmetry of a system as well as the valley and topological properties \cite{re1,re2,re3,re4,re5,re6,re7,fe,fe1,fe2,fe3,fe4,fe5}. For example, in monolayer $\mathrm{Fe_2Br_2}$, when the magnetic orientation is out-of-plane, the hot spots in the Berry curvature are around four gapped
Dirac cone with the same signs, leading to Chern number $C$=2; while for in-plane magnetization, two of four main hot spots in the Berry
curvature have the opposite sign with the other two, giving rise to a vanshing Chern number \cite{fe4}.
For our proposed system, an out-of-plane magnetic orientation is needed, and the magnetic orientation can be determined by magnetic anisotropy energy (MAE). By GGA+$U$+SOC method, the MAE can be calculated as $E_{MAE}=E^{||}_{SOC}-E^{\perp}_{SOC}$ where $||$ and $\perp$ denotes
the in-plane and out-of-plane spin orientation.
The MAE is 451$\mathrm{\mu eV}$/Fe, and the positive value indicates  the out-of-plane easy magnetization axis of $\mathrm{Fe_2BrMgP}$, which confirms our proposed design principles. The total magnetic moment per unit cell is strictly 0.00 $\mu_B$, and the magnetic moments of bottom/top Fe atoms are 3.06  $\mu_B$/-3.06 $\mu_B$.
\begin{figure}
  \includegraphics[width=8cm]{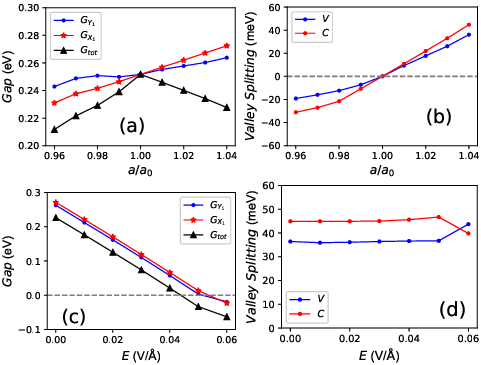}
  \caption{\textbf{Strain and electric field  modulation.} For $\mathrm{Fe_2BrMgP}$, the related bandgaps including the global gap [$G_{tot}$] and gaps of $Y_1$ and $X_1$ valleys [$G_{Y_1}$ and $G_{X_1}$] (a,c), and  valley splitting  for both valence [$V$] and condition [$C$] bands (b,d) as a function of $a/a_0$ (a,b) and $E$ (c,d) with $a/a_0$=1.04. }\label{gap}
\end{figure}

The calculated energy band structures of $\mathrm{Fe_2BrMgP}$ without and with SOC are shown in Fig.\ref{st} (c) and (d), respectively.
When neglecting SOC, two pairs of band-crossing points occurs near the Fermi level along $\Gamma$-X and $\Gamma$-Y lines.
With SOC, a Dirac gap of 252 meV is introduced, yielding equivalent valleys along the  $\Gamma$-X and $\Gamma$-Y lines due to $C_4$ rotation symmetry. The corresponding $k$ points in the momentum space are marked by $X_1$ and $Y_1$ without valley splitting ($\Delta E_C=E_{X_1}^C-E_{Y_1}^C$ and $\Delta E_V=E_{X_1}^V-E_{Y_1}^V$) for both conduction and valence bands. Because of the $\mathcal{PT}$ symmetry, the bands of $\mathrm{Fe_2BrMgP}$  are spin degenerate both without and with SOC.

Because of the $\mathcal{PT}$ symmetry, the Berry curvature of $\mathrm{Fe_2BrMgP}$ vanishes.  However, each layer breaks
the $\mathcal{PT}$ symmetry locally, and such layer-specific symmetr breaking leads to a non-vanishing layer-locked hidden Berry curvature. Here the berry curvatures of spin-up and spin-down channels are non-zero [Fig.\ref{t1} (a) and (b)]. The Berry curvatures are
opposite for spin-up and spin-down channels around $Y_1$ and $X_1$ valleys, respectively.
In the presence of a longitudinal in-plane electric field $E_{\parallel}$, the spin-up and spin-down electrons from  $Y_1$ and $X_1$ valleys will
accumulate along opposite sides of the top and bottom Fe layers, resulting
in layer-spin hall effect [Fig.\ref{fig1} (d)].

\textcolor{blue}{\textbf{Uniaxial strain induces valley polarization.--}} To induce valley polarization in $\mathrm{Fe_2BrMgP}$, an uniaxial strain along $x$ or $y$ direction is applied which reduces $C_4$ to $C_2$ symmetry. The valleys along the $\Gamma$-X and $\Gamma$-Y lines become inequivalent, thus giving rise to valley polarization.
We use  $a/a_0$ (0.96 to 1.04) to simulate a uniaxial strain along $x$  direction, and the lattice constants $b$ along $y$ direction is optimized. The in-plane Young's modulus $C_{2D}(\theta)$ as a function of the angle $\theta$ relative to
the $x$ direction is plotted in FIG. S1 of the Supplementary Materials. The obtained $C_{2D}$ along $x$ direction is 86 $\mathrm{Nm^{-1}}$. This is  smaller than those of graphene ($\sim 340\pm 40$ Nm$^{-1}$) and MoS$_2$ ($\sim 126.2$ Nm$^{-1}$) \cite{q5-1,q5-1-1}, which indicates the better mechanical flexibility of $\mathrm{Fe_2BrMgP}$, thus favoring the experimentally realization of valley polarization by strain.
The strained $\mathrm{Fe_2BrMgP}$ remains in the A-type AFM ground state with out-of-plane magnetic anisotropy within the considered strain range (see  FIG. S2 of the Supplementary Materials).

\begin{figure}
  \includegraphics[width=8cm]{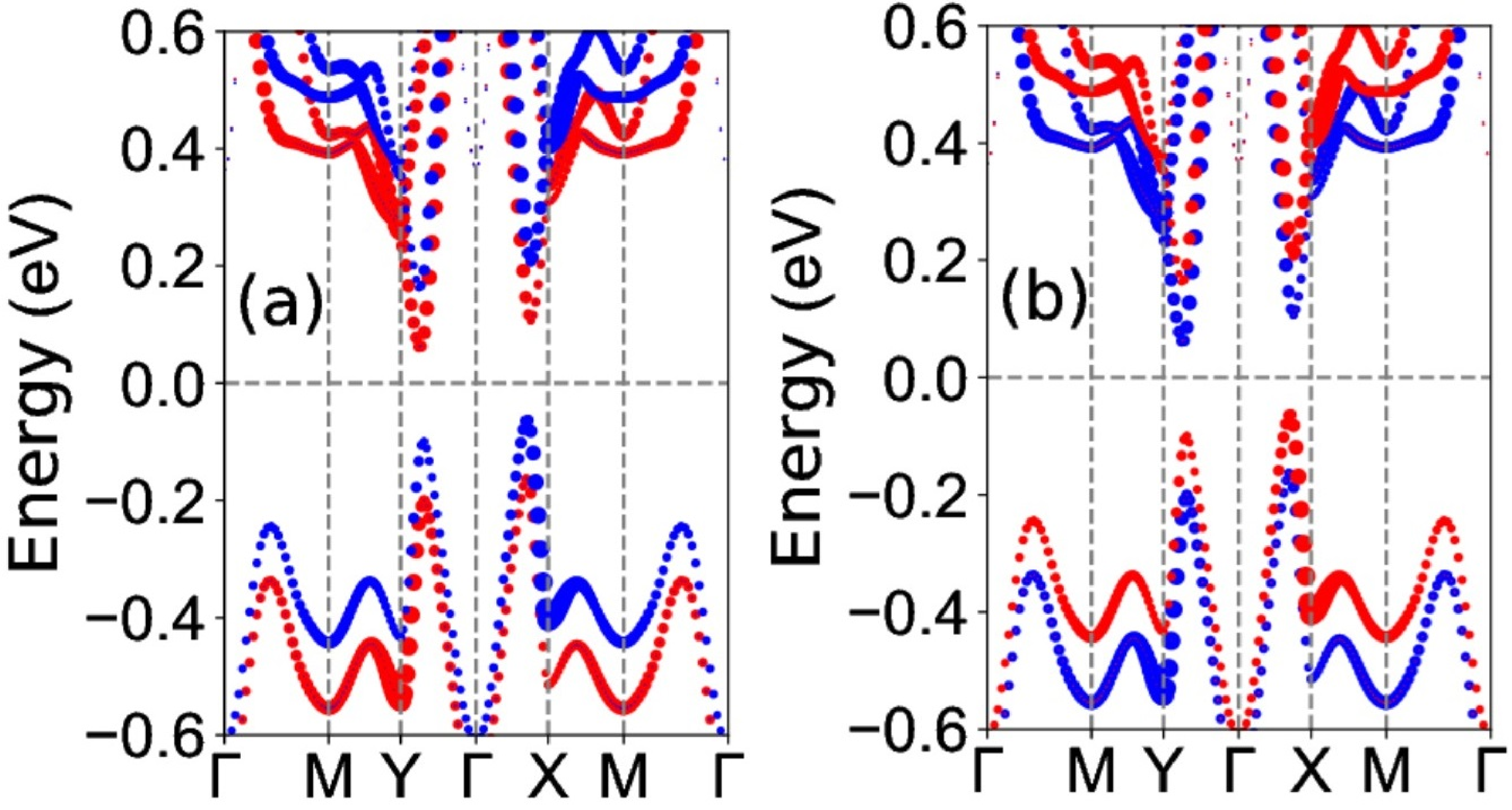}
\caption{\textbf{Sign-reversible layer-locked anomalous valley Hall effect.} For  $\mathrm{Fe_2BrMgP}$ with  $a/a_0$=1.04, the  spin-resolved energy band structures for $E$=+0.02$\mathrm{V/{\AA}}$ (a) and $E$=-0.02$\mathrm{V/{\AA}}$ (b). The spin-up
and spin-down channels are depicted in blue and red.}\label{pro}
\end{figure}

The electronic band structures of strained $\mathrm{Fe_2BrMgP}$ calculated using GGA+SOC are plotted in FIG. S3 of the Supplementary Materials. The evolution of the bandgap and the valley splitting ($\Delta E_V$ and $\Delta E_C$) for both valence and conduction bands as a function of $a/a_0$ are plotted in Fig.\ref{gap} (a) and (b). The uniaxial strain induces  valley polarization for both conduction and valence bands, and the valley polarization can be switched between $X_1$ and $Y_1$ valleys with strain transiting from compressive to tensile cases.
For common hexagonal FV systems, the valley polarization can  be reversed by magnetic field \cite{re1,re2,re3,re4,re5,re6}. Therefore, the uniaxial strain can be regarded as a pseudo-magnetic field for tetragonal system \cite{re7}.
For $a/a_0$=0.96/1.04, the corresponding valley splitting are -19  (-31) meV/36 (45) meV for valence (conduction) band, which are close to or  larger than the thermal energy of room temperature (25 meV).
Using $a/a_0$=1.04 as a representative, the Berry curvatures of the spin-up and spin-down channels are plotted in FIG. S4 of the Supplementary Materials. The Berry curvatures are
opposite for spin-up and spin-down channels around $Y_1$ and $X_1$ valleys.
In this case, a longitudinal in-plane electric field $E_{\parallel}$ can thus lead to the accumulation of the spin-up and spin-down electrons from only $X_1$  valleys along the opposite sides of top and bottom Fe layers, resulting in  valley layer-spin hall effect as illustrated in Fig.\ref{fig1}(d).

\textcolor{blue}{\textbf{Electric field induces spin splitting.-}} An out-of-plane  electric field  can  break the $\mathcal{PT}$ symmetry to lift the spin degeneracy of the valleys.
Taking strained $\mathrm{Fe_2BrMgP}$ with $a/a_0$=1.04 as an example, the  electric filed  ($E$)  effects  on the electronic structures  are investigated. The difference between  $+E$ and $-E$ is that the spin-splitting order is reversed.
According to FIG. S5 of the Supplementary Materials, the  ground state of $\mathrm{Fe_2BrMgP}$ rmains in the A-type AFM ordering with out-of-plane magnetic anisotropy within the considered $E$ range.
The energy band structures of $\mathrm{Fe_2BrMgP}$ by using GGA+SOC at representative $E$ are plotted in FIG. S6 of the Supplementary Materials, and  the  spin-resolved energy band structures at $E$=$\pm$0.02$\mathrm{V/{\AA}}$ are shown in Fig.\ref{pro}.
The evolutions of related energy band gap  and the valley splitting for both valence and condition bands  as a function of $E$ are plotted in Fig.\ref{gap} (c) and (d).

The spin splitting induced by out-of-plane electric field, and spin-polarization reversal via reversing the direction of electric field can be observed in Fig.\ref{pro}. The electric filed can maintain valley splitting amplitude, and induce a semiconductor-metal phase transition (see Fig.\ref{gap} (c) and (d)). The sizes  of spin splitting at $X_1$ and $Y_1$ valleys for both conduction and valence bands are plotted in FIG. S7 of the Supplementary Materials, which meets the layer-dependent electrostatic potential $\varpropto$ $eEd$ (The sizes  of spin splitting can be calculated by $eEd$).
Using $E$=0.02$\mathrm{V/{\AA}}$ as an example, the Berry curvatures calculations in Fig.\ref{t1} (c) and (d) show that the Berry curvatures of spin-up and spin-down channels around $Y_1$ and $X_1$  valleys are
opposite.
In the presence of a longitudinal in-plane electric field $E_{\parallel}$, the spin-up electrons from only  $X_1$ valley
accumulate along one side of bottom Fe layer, which leads to the layer-locked anomalous valley Hall effect of Fig.\ref{fig1} (d).
When the direction of electric field is reversed,  the spin-down electrons from only $X_1$  valley accumulate along the other side of top Fe layer, leading to a electric field-effect induced sign reversal of the AVHE previously predicted in FV-FM system \cite{zhou2021}.

\begin{figure}
  \includegraphics[width=8cm]{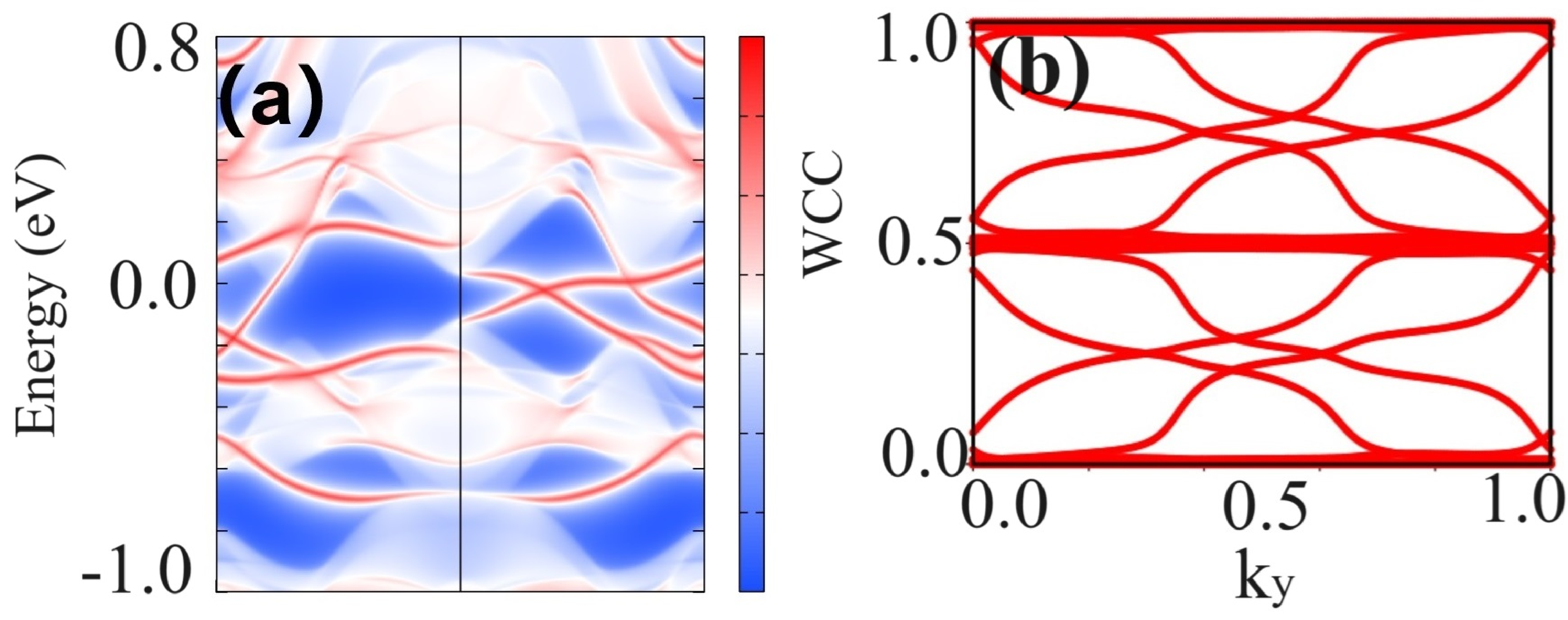}
\caption{\textbf{Topological properties of Fe$_2$BrMgP.} For  $\mathrm{Fe_2BrMgP}$ with  $a/a_0$=1.04 at $E$=+0.02$\mathrm{V/{\AA}}$: (a) the edge states  along the [100] direction; (b) the evolution of WCCs  along $k_y$. }\label{top}
\end{figure}

\textcolor{blue}{\textbf{Valley-polarized quantum spin Hall insulator.-}} The $\mathrm{Fe_2BrMgP}$ has been predicted to be an  AFM  quantum spin Hall insulator (QSHI) with high spin Chern numbers, as confirmed by the gapless edge states and the topological invariant spin Chern
numbers ($C_s$) \cite{x1}. By applying uniaxial strain and electric field simultaneously, the VQSHI can be achieved in $\mathrm{Fe_2BrMgP}$, which  combines AVHE and QSHE in one material,  providing a path towards integrating valleytronics, topological quantum
effects and spintronics in a single system. To confirm this aspect, for  $\mathrm{Fe_2BrMgP}$ with $a/a_0$=1.04 at $E$=+0.02$\mathrm{V/{\AA}}$, the edge states along the [100] direction and the evolution of the Wannier charge centers (WCCs)  along $k_y$ are plotted in Fig.\ref{top}.
Based on  the evolution of WCCs,  the spin Chern number $|C_s|$ is 2, which is further determined by two pairs of gapless edge states with opposite chiralities appearing  in the bulk gap. Therefore, VQSHI can indeed be realized in $\mathrm{Fe_2BrMgP}$.

\textcolor{blue}{\textbf{Conclusion.--}} In summary, we propose a paradigm for achieving anomalous valley Hall effect in AFM tetragonal monolayers by external field and strain engineering.
The proposed concept is confirmed by a prototype monolayer   $\mathrm{Fe_2BrMgP}$ using first-principles calculations.
Uniaxial strain induces valley polarization by breaking $C_4$ symmetry, and an out-of-plane electric field gives rise to spin splitting via layer-dependent electrostatic potential. The concept of VQSHI is demonstrated, which is similar to VQAHI. Our analysis can be readily extended to the broader family of $\mathrm{Fe_2XY}$ (X=or$\neq$Y=Cl, Br and I) and $\mathrm{Li_2Fe_2XY}$ (X=or$\neq$Y=S, Se and Te) bilayers as they share
the same Fe-dominated low-energy states as $\mathrm{Fe_2BrMgP}$.
Our results reveal a route towards
energy-saving and fast-operating spintronic-valleytronic devices based on 2D antiferromagnetic materials.

\begin{acknowledgments}
We are grateful to Shanxi Supercomputing Center of China, and the calculations were performed on TianHe-2. Y.S.A. is supported by the Singapore Ministry of Education Academic Research Fund Tier 2 (Award No. MOE-T2EP50221-0019).
\end{acknowledgments}

\end{document}